\documentclass[apj]{emulateapj}
\usepackage{amsmath}
\usepackage{amssymb}
\begin{document}

\newcommand\unit[1]{\,{\rm #1}}
\newcommand{\newtext}[1]{{\bf #1}}
\newcommand{\optional}[1]{}

\newcommand {\peryr}{$\rm{yr}^{-1}$}
\newcommand {\permyr}{$\rm{Myr}^{-1}$}
\newcommand{\nsns}{NS$-$NS\ }
\newcommand{\nswd}{NS$-$WD\ }
\newcommand{\wdns}{WD$-$NS\ }
\newcommand{\tlife}{\tau_{\rm life}}
\newcommand{\tmrg}{$\tau_{\rm mrg}$}
\newcommand{\td}{$\tau_{\rm d}$}
\newcommand{\tc}{$\tau_{\rm c}$}
\newcommand{\tsd}{$\tau_{\rm sd}$}
\newcommand{\rate}{$\cal R$}
\newcommand{\rtot}{${\cal R}_{\rm tot}$}
\newcommand{\rpeak}{${\cal R}_{\rm peak}$}
\newcommand{\rdet}{${\cal R}_{\rm det}$}
\newcommand{\ntot}{$N_{\rm tot}$}
\newcommand{\npsr}{N_{\rm PSR}}
\newcommand{\nobs}{$N_{\rm obs}$}
\newcommand{\nmean}{$<N_{\rm obs}>$}
\newcommand{\solarM}{M$_{\rm \odot}$}
\newcommand{\chmass}{$\cal M$}
\newcommand{\fmin}{$f_{\rm min}$}
\newcommand{\fmax}{$f_{\rm max}$}

\title{Mapping population synthesis event rates on model parameters
  II:  Convergence and accuracy of multidimensional fits}
\author{R.\ O'Shaughnessy,  V.\ Kalogera, \& K.\ Belczynski}

\begin{abstract}
Binary population synthesis calculations and associated predictions,
especially event rates, are known to depend on a significant number of
input model parameters with different degrees of sensitivity. At the
same time, for systems with relatively low formation rates, such
simulations are heavily computationally demanding and therefore the
needed explorations of the high-dimensional parameter space require
major -- often prohibitive -- computational resources. In the present
study, 
to better understand several key  event rates involving binary
evolution and binaries with two compact objects in Milky
Way-like galaxies and to provide ways of reducing the computational costs of complete parameter space explorations: (i)  we perform a methodical parameter study
of the  \emph{StarTrack} population synthesis code ; and (ii) we
develop a formalism and methodology for the derivation of {\em
  multi-dimensional fits} for event rates.   We significantly
generalize our earlier study, and we  focus on ways of thoroughly
assessing the accuracy of the fits.   We anticipate
that the efficient tools developed here can be applied 
in lieu of large-scale population calculations and 
 will facilitate the exploration of the
dependence of rate predictions on a wide range binary evolution
parameters. Such explorations can then allow  the derivation of
constraints on these  parameters, given empirical rate constraints and accounting for
fitting errors.
Here we describe in detail the principles and practice behind
constructing these fits, estimating their accuracy, and
comparing them with observations in a manner that accounts for their errors.
\end{abstract}

\maketitle

\section{Introduction}
Models for binary stellar evolution and population syntheses are 
necessary to provide quantitative theoretical predictions for the relative
likelihood of assorted events involving the evolution of binary
stars.  The resulting predictions are particularly critical when no empirical
estimates exist for topics of immediate astrophysical interest, such
as mergers of  double-compact object (DCO) through the emission
of gravitational waves.  The most practical and widely applied
binary population synthesis codes available in the community --  such as the
\emph{StarTrack}  code described in
\citet{StarTrack} and significantly updated in \citet{StarTrack2};
the BSE code, described in \cite{2002MNRAS.329..897H}; the SeBa code
described in \cite{1996AA...309..179P}; and the StarFaster code
described in \cite{1998ApJ...496..333F}  --  rely on a large
set of fairly simple parameterized rules to characterize many complex
and often ill-understood physical
processes.  Unfortunately, current binary population synthesis 
codes are greatly and often forbiddingly computationally demanding (depending  on their level of
sophistication): even with substantial simplifications,  
exploring the entire parameter space 
relevant to the simulations
 is beyond present-day
computational capability.  

However,  observational information can provide us with constraints
that help us improve our understanding of massive binary evolution.
For example, pulsar searches continue to discover and refine
observations of isolated pulsars and new binary pulsar systems;
e.g., see \citet{lrr-2005-7} for a review. 
Specifically, the samples of binary pulsars with neutron star and
relatively massive white dwarf companions have been used for a
statistical derivation of empirical rate estimates for their formation
(most recently see 
\citet{Chunglee-nsns-1}, \citet{Chunglee-wdns-1},  \citet{Chunglee-wdns-proceedings}
and references therein).  
Additionally, many ground-based gravitational wave detectors now operating at
or near design sensitivity (i.e., LIGO, GEO, TAMA) are 
designed to detect the late stages of double compact object (DCO) 
inspiral and merger.  Based only on early-stage data, these
instruments have already provided conservative upper limits to certain
DCO merger rates \citep[see,e.g.][]{LIGOS2nsns,LIGOS2macho}.  With
LIGO now very close to design sensitivity, a year of LIGO  data
could definitively exclude (and even possibly confirm) the most  optimistic theoretical
predictions for BH-BH merger rates \citep[see, e.g.][for a range of
BH-BH merger rates arising from binary evolution in Milky Way-like
galaxies]{PSutility}.  Thus, gravitational-wave based upper limits
(and, eventually, detections)  will shortly provide constraints on
theoretical models of DCO formation.

Faced with the availability of empirical rate constraints, and yet at first
unable to quantitatively impose them on population synthesis
predictions 
because of the extremely limited exploration of the
multi-dimenional parameter space,  
\citet{PSutility} realized that any single unambiguous population
synthesis  prediction could be sampled loosely and then fit
over the most sensitive population synthesis parameters.  
In the same study, we also presented a technique to accelerate the
synthesis code used (\emph{StarTrack}) to study a  single
target subpopulation, which we called ``partitioning'':
we used  experience gained from prior simulations to reject binary
parameters highly unlikely to produce the current event of interest.
\citet{PSconstraints} first applied these early fits to allow a direct
comparison between  \emph{StarTrack}-produced population synthesis
predictions and the observed formation rate for  NS-NS binaries.  
Though only a  small fraction ($2\%$) of \emph{StarTrack}
models appeared 
consistent with the constraints, conceptual challenges with
seven-dimensional visualization prevented \citet{PSutility} from
clearly describing the constraint-satisfying region.
A forthcoming paper, \citet{PSmoreconstraints} will significantly
extend that earlier preliminary analysis, adding significantly more
observational constraints as well as a clearer investigation of the
constraint-satisfying models.

In this paper, we extend and generalize the analysis of
\citet{PSutility}, and we
present a  thorough 
discussion of our much-updated and vastly larger population synthesis
archive (\S\ref{sec:ps:archive}) and particularly of the fitting methods we employ to extract
predictions and assess their uncertainties and quality
 (\S\ref{sec:fit}).   
Because several imporant  and yet not immediately obvious pitfalls
must be identified avoided  when constructing, testing, and applying
fits, in this paper we provide a thorough and pedagogical
discussion of our fitting method  and philosophy. 
For example, we explain how systematic fitting errors that were
ignored in \cite{PSconstraints} can
limit our ability to constrain the  merger rates in the Milky Way.
 In a companion
paper  \citep{PSmoreconstraints} we
explore the astrophysical applications of these fits, emphasizing
their use in deriving empirical constraints on population syntheses
and rate predictions.

Though we develop the fitting formulation specifically for the results of the \emph{StarTrack}
code, the methods described in this paper can be
applied to any family of population synthesis simulations.  For this
reason, we survey the physics underlying  the \emph{StarTrack} model only in our
companion paper \cite{PSmoreconstraints}, in which these fits are used
to  cover the parameter space and apply
empirical rate constraints from supernovae and binary pulsars.

\section{Population synthesis archives}
\label{sec:ps:archive}
Population synthesis simulations can be extremely computationally
demanding: even though \emph{StarTrack} can fully evolve roughly $10^3$
binaries of interest\footnote{Specifically $m_1>4 M_\odot$; see the discussion
  below.} per CPU-hour with modern-day processors, because some double
compact objects form very infrequently (e.g., black holes, which occur
roughly once every $\sim 10^{-4}$ binaries evolved), a
representative sample of stellar systems often contains
$10^{4.5}-10^{6.5}$ binaries and requires hundreds of CPU hours to
complete.   Additionally, since population synthesis rate predictions
depend delicately on model parameters, the \emph{computation time
  needed} to build up a sufficiently representative collection of
stellar systems -- one where some event of interest occurs many times -- 
varies considerably depending on astrophysical assumptions.    
Given the prohibitive computational demands of a brute-force approach, we
took  advantage of several simplifications originally
developed in \citet{PSutility} to assemble our archive of roughly 
3000 population synthesis simulations, upon which our fits
our predicated.
 In this section we
briefly describe how those archives were generated and how we identify
and extract event rates for several processes of interest.



\subsection{\emph{StarTrack} population synthesis code}
We estimate formation and merger rates for several classes of double
compact objects using the 
\emph{StarTrack} code first developed by \citet{StarTrack}
[hereafter BKB] and recently significantly updated and tested as described in detail in
\citet{StarTrack2}. 
Like other population synthesis codes, \emph{StarTrack} evolves
some number $N$ binaries from their birth (drawn randomly from specified birth distributions) to the present, tracking the stellar and
binary parameters.  
Because binaries without any high-mass components cannot produce black
holes or neutron stars, we configured \emph{StarTrack} to simulate
only those binaries whose heaviest initial component mass $m_1$ was
greater than $4 M_\odot$.  In effect, the small number $N$ of binaries
simulated mimic the results of a much larger simulation of size
$N_{eff}$ in which $m_1$ can take on any value from the hydrogren
burning limit ($0.08 M_\odot$) to the maximum initial stellar mass we allow
($150 M_\odot$):
\begin{eqnarray}
N_{eff}& =& N/\int_{4 M_\odot}^{150 M_\odot} dm \; \phi(m) \simeq   131 N
\end{eqnarray}
based on a broken Kroupa initial mass function $\phi(m) \propto m^{-1.3}$ if $ m\in[0.08, 0.5] M_\odot$, $ \propto
m^{-2.2}$ if $ m\in[0.5, 1] M_\odot$, and $ \propto m^{-2.7}$ if $ m > 1 M_\odot $.
In turn, a simulation of $N_{eff}$ stellar systems can be scaled up to
represent the $N_g$ stellar systems in the Milky Way.  The number of
stellar systems $N_g$ in the Milky Way (and thus the normalization of
population synthesis simulations) can be estimated in many ways,
depending on the observational inputs used; for the purposes of this paper, we
estimate $N_g$ from the average number of binary systems formed
through steady star formation over $T=10 \unit{Gyr}$ of star
formation at $\dot{M}
 \simeq 3.5 M_\odot \unit{yr}^{-1}$ \citep{RanaSFR91,ADM:Lac85}:
\begin{eqnarray}
N_g &=& \frac{\dot{M} T}{\left< m_1 +m_2\right>} = \frac{\dot{M} T}{ \left< m_1 \right>
     \left(1+ f_b \left<m_2/m_1\right>\right)} \\
  &\simeq& 7.6\times 10^{10} \left(\frac{T}{10 {\rm Gyr}}\right) 
        \frac{1}{1+f_b \left<m_2\right>/\left<m_1\right>} \nonumber
\end{eqnarray}
where $f_b$ is the binary fraction  and $\left<m_2\right>/\left<m_1\right>$ is the
ratio of average masses of the companion and primary, respectively.  
The proportionality constant $s\equiv N_{g}/N_{eff}$ needed to scale
the results of our simulations up to the universe is therefore
\begin{eqnarray}
\label{eq:def:s}
s &\equiv& N_g/N_{eff} \\
 &=& 5.9 \times 10^8 N^{-1}(T/10 {\rm Gyr}) \frac{1}{1+f_b
   [\left<m_2\right>/\left<m_1\right>]} \nonumber
\end{eqnarray}
Note that parameters of the population synthesis model such as
$f_b$ and $\left<m_2\right>/\left<m_1\right>$ influence the result at
best by $O(50\%)$.
These scaling relations allow us to estimate the merger rate implied
by simulations ($R$) via a surrogate ($\tilde{R}$) based on  the  number $n$ of
merger events seen in  simulation:
\begin{eqnarray}
\label{eq:def:Rtilde}
\tilde{R}&\equiv& s n/T = 7.6\times 10^{-3}\frac{n}{N} 
    \frac{\dot{M}}{\left< m_1 +m_2\right>}
   \\
 &\simeq&
  0.059 {\rm yr}^{-1} \frac{n}{N}\frac{1}{1+f_b [\left<m_2\right>/\left<m_1\right>]}
  \nonumber
\end{eqnarray}
[Unless otherwise noted, ``tilde'' quantities refer to \emph{estimates}
for the corresponding ``normal'' quantity for one individual
simulation, based on only information
from that one simulation.]

\optional{
Strictly speaking, this approach gives only the \emph{average} event rate.  The
present-day merger rate agrees with this quantity when most mergers occur
relatively  promptly  after their birth (relative to the age of the Milky Way
  and hence duration of the star formation phase, e.g., less than $100$ Myr).
  Some DCOs -- notably double BH binaries --  have substantial delays
  between birth and merger, introducing a strong time dependence to
  the merger rate.  The technique described above will significantly
  \emph{underestimate} these rates.  This point will be addressed in
  considerably more detail, both for the Milky Way and for a
  heterogeneous galaxy population in a forthcoming paper by
  \citet{PSgrbs-popsyn} and in greater detail in \citet{PSgrbs}.
}

Given unlimited computational resources, this  simulation could be
repeated many  times to  measure the  merger rate $R$ implied by this
model to any accuracy desired,  or 
equivalently measure the average number
$\mu$ of of merger events expected from this simulation
\begin{equation}
\label{eq:def:mu}
\mu \equiv R T/s \, .
\end{equation}
 In practice, each simulation is performed
once; therefore, the simulated number of merger events ($n$)  is only
statistically correlated to the average number  of events expected ($\mu$), with relative
probabilities $p(n|\mu)$ of any one simulation producing $n$ events given by the Poisson distribution 
\begin{equation}
\label{eq:def:poisson}
p(n|\mu) = \frac{\mu^n}{n!} e^{-\mu} \; .
\end{equation} 
On average our estimates of the  $\tilde{R} = s n/T$ and 
equivalently of the number of events 
$
\tilde{\mu} = n \; 
$
will agree with the true properties of the underlying simulation:
averaging over many trials, $\left<\tilde{R}\right>=R$ and 
and $\left<\tilde{\mu}\right>=\mu$.  However, an estimate 
based on any one specific simulation should differ from the mean by a characteristic
relative amount 
\begin{eqnarray}
\label{eq:def:samplingerror}
\sqrt{\left<(\log \tilde{R} - \log R)^2
  \right> }
&=&
\sqrt{\left<(\log \tilde{\mu} - \log \mu)^2
  \right>} \\ 
&\simeq&1/\sqrt{\mu}\ln(10) \nonumber 
\simeq 1/\sqrt{n}\ln(10)   \nonumber
\end{eqnarray}

\subsection{Parameters varied in archives}
Our extensive experience in modeling of binary compact objects with {\em StarTrack} clearly indicates that there are seven parameters that strongly
influence compact object merger rates (see e.g., \citet{StarTrack}):
the supernova kick distribution
(3 parameters $\sigma_{1,2}$ and $s$ describing a superposition  of two 
 independent maxwellians), the massive stellar wind strength $w$ (1), the 
common-envelope energy transfer efficiency $\alpha \lambda$ (1), the fraction of mass
accreted by the accretor in phases of non-conservative mass transfer $f_a$
(1), and the binary mass ratio distribution, as  described by a
negative power-law index $r$ (1).
We allow the dimensionless parameters
$\alpha\lambda,f_a$, $w$ and $s$ to run from 0 to 1; the dimensionless
$r$ can be between $0$ and $3$; and finally we vary the
dispersion of either component of a bimodal Maxwellian
$\sigma_1,\sigma_2$ from 0
to 1000 km/s.  
Additionally, motivated by recent observations suggesting  pulsars
with masses near $2 M_\odot$ \citep[see,e.g.][]{2005ApJ...634.1242N,psr:measurement:J0621+1002},
we assume the maximum neutron star mass to be $2.5
M_\odot$\footnote{Such a high neutron star mass
  converts many merging binaries we would otherwise interpret as BH-NS or even
  BH-BH binaries into merging NS-NS binaries, driving down the BH-NS
  and BH-BH rates significantly from the distributions shown in
  \cite{PSconstraints}, 
as also described in \S,4.
}.

\begin{figure}
\includegraphics[width=\columnwidth]{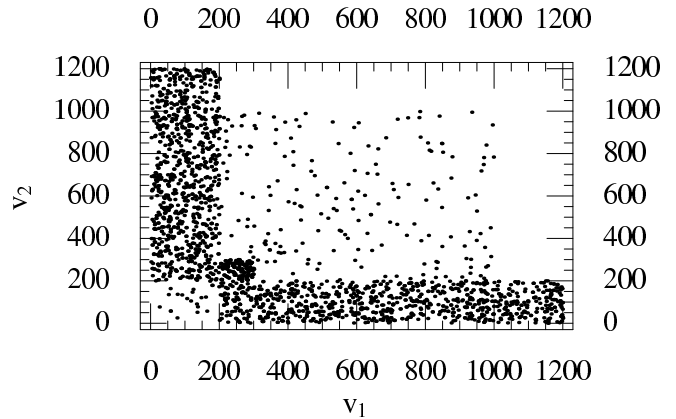}
\caption{\label{fig:popsyn:kicksampling} Scatter plot of the two
  bimodal kick velocity dispersions $\sigma_1$, $\sigma_2$ for the population synthesis
  archives used to evaluate the merger rate of NS-NS binaries [denoted
  NSNS(m)]. The strong bias is introduced by incorporating
  high-density and computationally expensive simulations
  from \citet{PSutility}
  and \citet{PSconstraints}. 
}
\end{figure}

To improve our statistics, we combine results from several different
databases of simulations.  The most extensive database samples
$\sigma_1\in[200,1200]$ and $\sigma_{2}\in[0,200]$ very densely and was
developed by \citet{PSutility} and \citet{PSconstraints}.  A
second archive, significantly less dense due to computational resource
limitations, allows both dispersions to run uniformly from 0 to 1000
km/s.  [Additional archives include, for example, a set chosen to better-sample kick
parameters that best correspond to observations of pulsar proper
motions \cite{Ar,HobbsKicks}.]
Consequently, as shown in Figure~\ref{fig:popsyn:kicksampling}, our
archived results do not uniformly sample the kick-related parameters
through this range.   
This irregular sampling has two effects.
First, having irregular sampling of the model parameters effectively
corresponds to imposing non-flat priors on these parameters and hence
biasing the resulting distribution function of merger rates that come
directly from the database of runs;  see  Appendix
\ref{ap:sampling} and in particular Figure
\ref{fig:popsyn:distributions}.
 Therefore it is even more
important to develop the fits, which then allow us uniform sampling of
the parameter space, and hence the derivation merger rate
distributions assuming flat priors, as our intention is. Second, 
%
%
certain kick combinations are relatively undersampled, which likely 
  plays a role in the relatively poor \emph{global} convergence of fits for physical
  parameters, as described in \S~\ref{sec:ps:predictions}.  Nonetheless,
  our sampling rather thoroughly explores the most physically likely
  regimes suggested by \citet{HobbsKicks} and \cite{Ar}.

\subsection{Event identification}
\label{sub:id}
Most events of physical interest are uniquely and fairly unambiguously
identifiable within the code.  Type II and Ib/c supernovae events are 
distinguished by the presence or absence of a hydrogen-rich envelope
at the supernova event.  We also record DCOs which
merge, so we can unambiguously determine the number of BH-BH, BH-NS,
NS-NS, and WD-NS merger events which occur in a simulation.  [These event
classes will be denoted BHBH(m), BHNS(m),  NSNS(m), and WDNS(m) for brevity
henceforth.]  Since the code also tracks binary eccentricity, for
example, we can identify those WD-NS binaries which end their
evolution with a non-zero eccentricy [denoted WDNS(e)].

When constructing our archived population synthesis results, we do
not record detailed information about the nature and amount
of any mass transfer onto the first-born NS.  We therefore cannot
determine the degree to which pulsars are recycled.   However, we do
record whether any mass transfer occurs.  Thus for the purposes of 
identifying a class of potentially recycled (``visible'') wide NS-NS
binaries [denoted NSNS(vw)], we assume any system undergoing \emph{non-CE} mass transfer
recycles 
its NS primary. 


\subsection{Practical Archive Generation and Resolution, with
  Partitions and Heterogeneous Targets}

The accuracy to which each population synthesis event rate prediction
is known, $\sim 1/\sqrt{n}$, is
uniquely set by the number of events $n$ seen within that simulation.
For this reason, \citet{PSutility} (i) designed their population synthesis
runs to continue until a \emph{fixed number of events were seen} and
(ii) used results only from such targeted simulations, where a minimum
number of events was guaranteed.  
In contrast in this study, given limited computational resources and a wide range of targets for which
predictions are needed, we extract
\emph{all possible information} from each simulation: whenever
possible, we make an estimate of each event rate of interest.  

However, it is important to note that most of our simulations employ
some degree of 
accelerating simplification which can \emph{bias} estimates.  To give the most
extreme example: the most accurate estimates for the BH-BH merger rate
come from population synthesis runs which evolve only a subset of
possible progenitor binaries by using \emph{partitions}.
This subset has been shown to include the progenitors of the vast
majority of BH-BH binaries, but very
few  progenitors of NS-NS binaries and other less massive DCOs
\citep[for more information, see][; we continue to employ the same
partition they devised]{PSutility}.  Similarly, the vast
majority of simulations used to study NS-NS event rates (i) use a
similar partition to reduce contamination from white dwarf binaries
and (ii) terminate the evolution of any binary immediately after any
WD forms.   These strong biases make data from these two types of
simulations inappropriate for use in, for example, estimating the WD-NS
merger rate.  For this reason, the number of population synthesis archives $N$
available to make predictions varies significantly across the various
target event types; see Table~\ref{tab:SimulationsAndFits}.  

Furthermore, the
accuracy each simulation can provide varies considerably: unlike
earlier studies,
the number of event samples $n$ in each archive is not guaranteed to
be greater than a minimum value.
However, usually we have at least one event for each event type:
$N_{s+}$, the number of unbiased population synthesis results
containing one or more events is usually very close to $N_s$, the
total number of unbiased population synthesis simulations available.
And in most cases a significant proportion of our sample contains
enough events to determine the rate to better than $30\%$ (i.e., $n\ge
10$).




\section{Fitting: Principles and Tests}

\label{sec:fit}
Given the prohibitive computational demands of direct population
synthesis simulations described in \S~\ref{sec:ps:archive}, we use fits based on archived results of  population
synthesis runs as a  surrogate for repeated detailed  simulations.
Confidence in our results is therefore tied intimately to confidence
in the quality of these
fits.  However, even low
order fits in seven dimensions involve many parameters: to fit any
nontrivial function, we \emph{must} fit roughly a handful of data points per parameter.
To build confidence, we must  show that the fit order chosen
adequately describes the data without overfitting.  More delicately,
since these fits are used in the derivation of empirical constraints
on the population synthesis models in our companion paper
\citep{PSmoreconstraints}, we must also be able to show that the key
end product, the "constraint-satisfying model region"  does not depend
sensitively on the 
fit details or on random accidents in the data (i.e., any different
Monte Carlo realization of our simulations should yield the same
result).  
 Finally, we note that our companion paper \citep{PSmoreconstraints}
will impose \emph{four}
 independent constraints.  In order to have more than $90\%$ confidence
 that all models that satisfy all constraints are indeed inside the
 intersection of the four constraint-satisfying regions, we must at
 a minimum  show that
 each individual  constraint-satisfying prediction contains more than
 $0.90^{1/4}\simeq  0.974$ of 
 the models that truly satisfy that constraint.

\subsection{One-Dimensional Model}

Though we intend to build confidence explicitly in our ability to make
predictions  on the basis of seven-dimensional fits, some of the required principles,  notation, and tools are best illustrated through a one-dimensional example.
Thus in this section we explore an arbitrarily-chosen but known 
  ``population synthesis'' model.  Depending on 
one real parameter $x$ with $0\le x \le 1$, this  model
on average will produce $\mu(x)$ merger events out of $N=10^5$
binaries, with $\mu(x)$ chosen  arbitrarily as
\begin{eqnarray}
\label{eq:OneD:mu}
\mu(x)&\equiv& 10^{M(x)}=5+20 x^{2}  \; . 
\end{eqnarray}
Specifically, in this model we assume: all stars are
born in binaries  
($f_b=1$); that the smaller companion star has mass randomly
distributed from the hydrogen burning limit to the primary's mass,
corresponding to 
$\left<m_2\right>/\left<m_1\right>=0.5$; and  that star formation extends
over an interval $T=10\unit{Gyr}$.  Under these specific circumstances, as 
discussed in \S\ref{sec:ps:archive}
[Eqs. (\ref{eq:def:mu},\ref{eq:def:s})],
a model which produces $\mu(x)$
events  on average out of $10^5$ binaries corresponds to a merger rate
\begin{eqnarray}
R(x) 
 &=& s \mu(x) /T \\
   &\simeq& 4\times 10^{-7} \mu(x) \unit{yr}^{-1} \; . 
\end{eqnarray}
An individual simulation of this model will produce some number $n$ of events, with $n$
statistically related to $\mu(x)$ via the Poisson distribution [Eq. (\ref{eq:def:poisson})].

\begin{figure}
\includegraphics[width=\columnwidth]{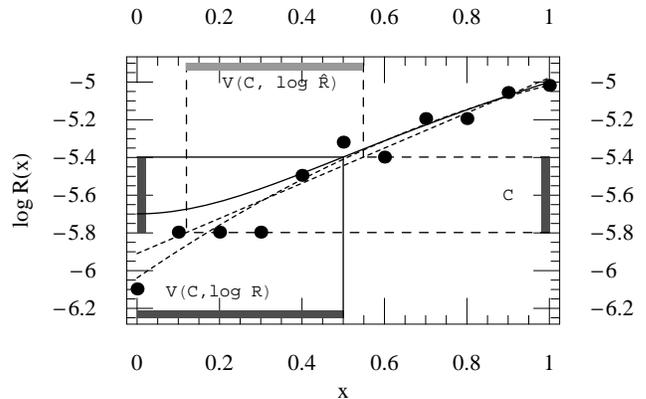}
\caption{\label{fig:Motivation:DataScatterPlot}  Illustration of
  how small errors in fits can produce large uncertainties in the
  constraint-satisfying region,
  using a one-dimensional model. 
  Solid line: the ``true'' function 
  $\log R(x)=\log \mu(x) s/T$ [Eq. (\ref{eq:OneD:mu})], expressing the log of the average
  number of events expected for any $x$.  Points: results of ``experiments''
  for eleven values of the parameter $x$. Dotted lines: linear and quadratic
  least-squares fits to the data.  Dark shaded boxes: (y-axis) the
  ``constraint'' interval $C=[-5.8,-5.4]$; (x-axis) the range of $x$ which
  truly satisfies the constraints [i.e., all $x$ so $ \log R (x)\in C$; this
  region is denoted  $V(C,\log R)$ in the notation of this paper].
  Light shaded box: the interval \emph{predicted} to satisfy the
  constraints, based on a linear fit [denoted $V(C,\log \hat{R}_1)$ in the
  notation of this paper].
}
\end{figure}

To estimate the merger rate as a function of $x$, we perform several simulations ($N_s=11$)
using different parameter values $x=x_\alpha = (\alpha-1)/10$ for
index values $\alpha = 1 \ldots 11$, 
obtaining results $n_\alpha$ and thus estimates $\tilde{R}_\alpha = s n_\alpha /T$ as
shown  in  Figure \ref{fig:Motivation:DataScatterPlot}.
To better fit the merger rate over the many orders of magnitude that
appear in practice, we choose our fit
$\hat{R}(x)=10^{\hat{M}(x)} s/T$ to minimize 
the \emph{logarithmic} difference between it and simulations,
weighted by the relative statistical uncertainties of each simulation,  $1/n (\ln 10)^2$ from Eq. (\ref{eq:def:samplingerror}).
[Here and henceforth we use ``hat'' symbols to denote our
fit quantities; unlike
``tilde'' quantities, which estimate properties of an individual simulation, the ``hatted'' quantities generally depend on
the results of all simulations  simultaneously.]
Specifically, to generate a $q$-th order polynomial fit
$\log \hat{R}_q(x)$ for the logarithm of the merger  rate, we choose
the 
coefficients of the polynomial $\hat{M}_q(x)=\log \hat{R}_q(x)T/s$ to
minimize the ``average square deviation divided by characterist deviation,  per degree of freedom''
\begin{eqnarray}
\label{eq:physical:diagnostic:chi2q}
\chi^2_q &\equiv &
\frac{1}{N_s-N_{\le q}} \sum_{\alpha=1}^{N_s} [\log_{10}(\hat{R}_q(x_\alpha)/\tilde{R}_\alpha)]^2 n_\alpha (\ln 10)^2
\end{eqnarray}
where $N_{\le q}$ is the number of $q$th-order basis polynomials ($q+1$, in
one dimension).   
The fit
 which minimizes $\chi^2_q$ is  the (gaussian) maximum-likelihood
 estimator for $R(x)$,
the best single estimate possible.%
\footnote{We have also appled a
maximum-likelihood estimate based on Poisson rather than (approximate)
gaussian errors.  For the problems explored here, for which many
merger events are usually available, we see no significant difference between
the results of this more statistically appropriate method and the
conventional and pedagogically far simpler gaussian maximum-likelihood
method. 
}
 Figure  \ref{fig:Motivation:DataScatterPlot} shows
the results of this minimization for the linear ($q=1$) and quadratic
($q=2$) fits:
\begin{eqnarray}
\log \hat{R}_1 &=& \log(s/T)+ 0.49 + 0.93 x \\
\log \hat{R}_2 &=& \log(s/T)+ 0.35 +1.49 x - 0.547 x^2 
\end{eqnarray}

Both performing simulations and fitting
 introduce errors,
which  we can quantify
by the mean square deviation between the exact merger rate $\log R$ and its fit $\log \hat{R}_q$:
\begin{eqnarray}
\label{eq:def:Iq}
  I_q &=&
 \left[\int dx |\log (\hat{R}_q(x)/R(x))|^2\right]^{1/2}  
\end{eqnarray}
For example, the value of $I_1$ is $0.075$.
Roughly speaking, $I_q$ estimates the rms uncertainty in the fit:
$\hat{R}_q$ should lie within a factor $\sim 10^{\pm I_q}$ of $R(x)$.
For example, $R_1$  should be accurate  to
within a factor roughly $10^{\pm 0.075}\simeq 1.2^{\pm 1}$, as
confirmed by Fig. \ref{fig:Motivation:DataScatterPlot}.\footnote{Of
  course, regions where the fits perform significantly worse can
  exist; here, the fit differs from $\log R(x)$  by
  $\simeq 6.5 I_1$ near $x=0$.}
Similarly, the values of $x$ that correspond to a given merger rate
should be uncertain to roughly $O\left(I/(\partial \log R/\partial
  x)\right)\simeq 0.1$.

As noted previously, an extremely high level of confidence is needed
in any prediction of a constraint-satisfying region.  Let us use
$V(G,f)$ to denote the set of points that $f$ maps to an interval
$G=\{g_{\rm min}, g_{\rm max}\}$, so
$V(G,f)=\{x| f(x)\in G\}$; and  $C$ is an observation we demand our
simulation reproduce; then the fraction of  $V(C,\log R)$ -- the  population
synthesis models which \emph{truly} satisfy the constraint -- which
are inside  $V(C, \log \hat{R})$  -- the set of population synthesis
models which are naively \emph{predicted} to satisfy the constraint,
based on the fit and the original constraint -- is
given by  $r_+(C,\log R|C,\log\hat{R})$ where $r_+$ is generally
defined by
\begin{equation}
\label{eq:rplus}
r_+(A,\log R|B,\log\hat{R}) = \frac{|V(A \log R)\cap V(B \log \hat R)|}{|V(A \log R)|}
\end{equation}
for arbitrary intervals $A,B$; here  $|V|$ is the volume of
$V$.  In the example
shown in Figure
\ref{fig:Motivation:DataScatterPlot}, the fraction $r_+$ of the \emph{truly
  constraint-satisfying} area  that is \emph{predicted} to
satisfy the constraints is only $78\%$; if this constraint had to be
combined with four other constraints of similar quality, fewer than
$(78\%)^4\simeq 37\%$ of all constraint satisfying points would lie
inside the intersection of all four naive predictions.

To increase the fraction of constraint-satisfying points included in a
prediction -- in the notation of this paper, to find a volume $V$
which contains the ``naive'' prediction $V(C,\log \hat{R}_q)\subset V$ but
contains more of the truly constraint-satisfying points $V(C,\log R)$ --
the simplest option is simply to increase the size of
the ``constraint'' interval.  A larger constraint interval $C^*$
containing the original would automatically include more points in its
predicted region (now $V(C^*,\log \hat{R})$).  While this prediction
loses some of its 
\emph{reliability} (i.e., we cannot ensure that only constraint-satisfying
points lie in this region), we increase its \emph{robustness}: by a
good choice of $C^*$ we
can almost guarantee that all constraint-satisfying points are
included. 
Specifically, if we change the interval $C=\{c_{\rm min}, c_{\rm max}\}$
to the broader interval $C^*(I)=\{c_{\rm min}-I, c_{\rm max}+I\}$
-- that is, if we increase the constraint region by the size of the
characteristic error in the fit --  then
any point $x$ which truly satisfies the constraints (in our notation,
$x\in V(C,\log R)$) almost certainly lies within the
larger ``predicted'' volume $V(C^*,\log \hat{R})$.  Formally,
assuming the error at any point $|\delta \log R|$ is likely less than
$I$, we conclude that  $\log \hat{R}(x)=\delta \log R(x) +\log R(x) >
\delta \log R(x) +c_{\rm min} \gtrsim - I + c_{\rm min}$ and similarly that
$\log \hat{R}(x)\lesssim I+ c_{\rm max}$. 
 In the case shown in Figure
\ref{fig:Motivation:DataScatterPlot},
 this error-widened prediction
$V(C^*(I_1),  \log \hat{R}_1)$  includes
both a higher fraction of  constraint satisfying points ($r_+\simeq
92\%$)  and a higher fraction of points which are not constraint satisfying
($20\%$). 
\optional{ A marginally larger increase in the constraint interval
would recover \emph{almost all} the constraint-satisfying points,
while simultaneously producing a prediction which is predominantly
($>75\%$) constraint-satisfying.   And though individually quite weak,
several of these constraints combined can and do extremely
strongly limit the volume that contains all constraint-satisfying points.
}

\begin{figure}
\includegraphics[width=\columnwidth]{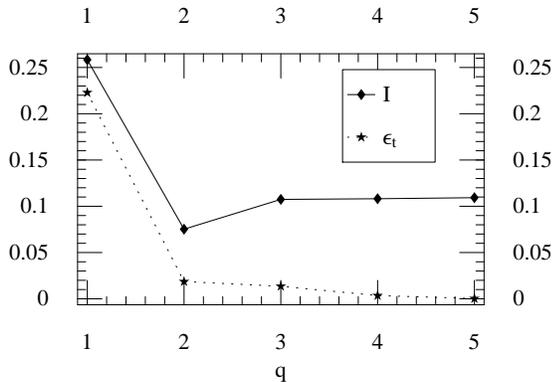}
\caption{\label{fig:popsyn:oneDconverge} Illustration of the balance
  between truncation and sampling error in a one-dimensional fit.}
\end{figure}

All these calculations, however, depend on the \emph{fit order}:
higher fit orders $q$ often allow more degrees of  
freedom with which to reproduce the true function $\log R(x)$ and
therefore estimate $V(C,\log R)$.  For
example, assuming arbitrarily small (constant) errors and arbitrarily dense
sampling $\vec{x}_\alpha$, the best possible $q$-th order
polynomial least-squares fit is the orthogonal projection $P_q \log R$ of $\log R$ onto the space
of $q$-th order polynomials, given in one dimension by
\begin{eqnarray}
\label{eq:project:oned}
(P_q \log R)(x) &=& \sum_{l=0}^q
    \phi_l(x) 
    \int_0^1 dy \phi_l(y) \log R(y) \nonumber
\end{eqnarray}
for $\phi_l(x)$ any set of orthonormal basis polynomials of degree
$\le l$ (e.g.,
$\phi_l(x)=\sqrt{2l+1}P_l(2x-1)$ where $P_l(x)$ are Legendre polynomials).  Therefore, the minimum possible error of a
$q$th-order polynomial fit is the magnitude $||P_{\perp q} \log R||$
of the difference $P_{\perp q}\log R\equiv  \log R - P_q \log R$
between $\log R$ and this best fit, where 
$||f||^2 = \int_0^1 dx |f(x)|^2$.  As the fit order
increases, this ``truncation error'' 
\begin{equation} 
\epsilon_{t,q}\equiv  ||P_{\perp q} \log R ||   
\end{equation}
 decreases dramatically, as shown in  Figure \ref{fig:popsyn:oneDconverge}.
%
However, the total error $I_q$ in the fit reflects
a balance between (i) decreasing the ``truncation error'' with more
degrees of freeom and (ii) increasing the ``sampling error'' as fewer data
points are available per degree of freedom.  For this reason, an
\emph{optimal fit order} exists, which provides the smallest error in
$R(x)$, indicated by a \emph{minimum} in $I_q$.  This best fit order will
provide the most reliable estimate of $V(C,\log R)$.

In principle, this one dimensional example contains all
the core concepts we will apply to fitting seven-dimensional
population synthesis results: the sampling uncertainty in  each sample
$n_\alpha$ of ``simulated merger events''; the fit and its
inaccuracies; the need to increase the size of the constraint interval
and thus the ``predicted'' constraint-satisfying region to include a
larger fraction of constraint-satisfying events; and a specific procedure for doing so,
based on the characteristic errors of the fit.  
%
In practice, however, we must substantially flesh out this outline
in two ways: 
(i) we must add a reliable way for estimating the fit error and
selecting the optimal fit order without using $I$, since its
definition in Eq. (\ref{eq:def:Iq}) uses the fit itself and requires exact, a
priori knowledge of $R(x)$ (the function being fitted) that is not available in practice; 
(ii) we
must extend the concepts developed here to seven dimensions, given the
dependence of our population simulations.

\subsection{Seven-Dimensional Analog}
\label{sec:sub:7dmodel}
By virtue of existing in seven dimensions, the full
population synthesis fitting problem is qualitatively different than
the
one-dimensional problem even with the same number of points and
characteristic error.  In the
one-dimensional case, the  many  densely packed and
relatively accurate  samples of $R$
permit very accurate fits to $\log R$; barring rapid variation in
$\log R$ with $x$,  one-dimensional fits can easily have accuracies considerably
exceeding the limiting accuracy of any individual measurement.  
On the other hand, in seven dimensions the same number of data points
are much more loosely spaced (geometrically, the characteristic
spacing is proportional to $ N^{-1/d}$) and must constrain many more degrees of
freedom at the same polynomial order, yielding much less accurate
fits.   
For this reason, rather than attempt to rescale the number 
and uncertainties of our population synthesis simulations to generate
a tractable one-dimensional analog, we instead demonstrate convergence
using a seven-dimensional model that best captures the relevant features of our
population synthesis data.

Explicitly, in this seven-dimensional toy model  (i) the logarithm of the simulation size $\log N$ is gaussian
distributed around $-4.5$ with an order of magnitude standard
deviation, omitting simulations smaller than $10^3$ and larger than
$10^7$; (ii) each set of  simulation parameters $\vec{x}_\alpha$ is
chosen by uniformly selecting seven random numbers  $0\le x^{k=1\ldots
  7}_\alpha\le 1$;  (iii) the number of mergers $n_\alpha$ observed in any
particular simulation parameterized by the seven parameters
$\vec{x}_\alpha$ is drawn from a Poisson distribution 
with mean $\mu(\vec{x}_\alpha)=\rho(\vec{x}_\alpha) N$, where the mean number of
mergers per binary $\rho(x)$ is defined by
\begin{eqnarray}
\label{eq:7d:mu}
 \rho(\vec{x})=\mu(\vec{x})/N &=& 10^{-4 x\cdot x/7-2.5} \;,
\end{eqnarray}
and thus (iv) implies a merger rate 
\begin{eqnarray} 
\label{eq:7d:R}
 R(\vec{x})  &=&\rho(\vec{x})(0.04 {\rm yr}^{-1}) 
\end{eqnarray}
(again assuming   $f_b=1$ and
$\left<m_2\right>/\left<m_1\right>=0.5$).  More specifically
still, to compare with the array of $O(2000)$ simulations of merging
NS-NS binaries, we generate a ``model archive'' of $N_s=2000$ randomly
realized analogs, each defined by their
parameter choices $\vec{x}_\alpha$, 
specific simulation sizes $N_\alpha$, 
observed merger numbers $n_\alpha$, 
and merger rates $\tilde{R}_\alpha =  0.04 {\rm yr}^{-1}(n_\alpha/N_\alpha)$.

\begin{figure}
 \includegraphics[width=\columnwidth]{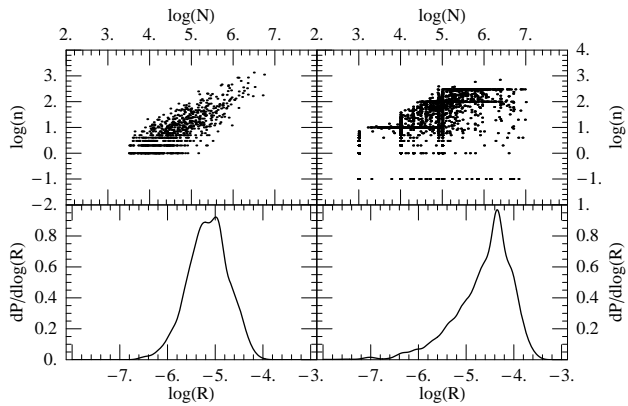}
\caption{\label{fig:analog} 
Comparison between our toy model (left) and simulations of merging NS-NS
binaries (right).  Top panels: scatter plot of the simulation size $N$
versus the number of merger events $n$ seen.  Bottom panels:
distribution of merger rates seen in simulation.  In the right panels,
no corrections are being made for the highly irregular patterns in
which $N$ and $n$ were chosen, and which introduces unavoidable biases
in the distribution of $\log(R)$.   
}
\end{figure}

This seven-dimensional model qualitatively reproduces the distribution of merging double
neutron star [NS-NS(m)] simulations in
$(n,N)$ and $R$ [Fig. \ref{fig:analog}].  Quantitatively, this toy model
is on average as accurate as our population synthesis simulations: the
root-mean-square relative sampling error  $\sigma_E$ (the ``expected''
standard deviation in the data, given the limiting error produced by
sampling at each point)
\begin{eqnarray}
\label{eq:physical:diagnostic:sigmaBetweenFitAndData}
\sigma^2_{E} &\equiv &\frac{1}{N_s}\sum_{\alpha=1}^{N_s} 
   \left[n_\alpha (\ln 10)^2\right]^{-1} 
\end{eqnarray}
of our population synthesis simulations $\sigma_E=0.25$ [see Table
\ref{tab:SimulationsAndFits}] agrees with the
corresponding value $\sigma_E=0.24$ [see Table
\ref{tab:analogConvergence}] for this seven-dimensional analog.
Furthermore, in both cases this average accuracy $\sigma_E$ is still slightly smaller than
the range in $\log R$, as characterized by $\sigma_{DD}$ (the
``data-data'' standard deviation, measuring the range of the distribution):
\begin{eqnarray}
\label{eq:physical:diagnostic:sigmaOfData}
\sigma_{DD}^2 &=&\frac{1}{N_s}
   \sum_\alpha\left[(\log \tilde{R}_\alpha)^2 - \left< \log \tilde{R}\right>^2\right]
\end{eqnarray}
 ($\sigma_{DD}=0.73$ for real simulations; $\sigma_{DD}=0.42$ for our toy model)
Finally, as can be seen by comparing  $N_s$, the
number of population synthesis simulations, to $N_{s}^+$, the number
of simulations with at least one event [Table
\ref{tab:analogConvergence}], a significant fraction $O(10\%-50\%)$ of population
synthesis simulations are totally unresolved; our seven-dimensional toy model has a
similar fraction ($20\%$) of unresolved simulations.

\subsection{Error  Estimates and Convergence}
Using this concrete and highly realistic seven-dimensional model
problem described in \S \ref{sec:sub:7dmodel}, we can not only conclusively demonstrate that fairly accurate fits are
possible for population syntheses, but we can also find a reliable estimate
of the rms error in the fit, an estimate that can be reliably applied
to surround all constraint-satisfying points inside a ``predicted''
volume.
Precisely as in the one-dimensional case, given a fixed archive of seven-dimensional simulations, 
we can estimate  $\log R$ with $\log \hat{R}_q$, the
unique polynomial  of order $\le q$ that minimizes $\chi^2_q$, using
any integer $q$ [Table \ref{tab:analogConvergence}].   
With more
degrees of freedom, higher-order fits can 
much more easily reproduce the data, so their mean square difference
from the data
$\sigma_{DF}$ 
\begin{eqnarray}
\label{eq:physical:diagnostic:sigmaFit}
\sigma_{DF}^2 &\equiv&  \frac{1}{N_s}
  \sum_\alpha \left[\log\hat{R}(\vec{x}_\alpha)  - \log \tilde{R}_\alpha \right]^2
\end{eqnarray}
(a ``data-fit'' comparison) decreases monotonically; see the fourth
column of Table \ref{tab:analogConvergence}.  
But as these high order polynomials increasingly fit every
sampling-induced flucuation in the data, high-order fits increasingly
deviate from the exact solution: the exact rms error $I_q$  is also
shown in Table \ref{tab:analogConvergence} [calculated by comparing
the fit to the known
seven-dimensional model $R(x)$ from Eqs. (\ref{eq:7d:R,eq:7d:mu})], showing
 $I_q$ has a minimum near an optimal fit order.  
Near this optimal fit order, the characteristic
fit error $I_q$ is substantially less than the characteristic range of
the data ($\sigma_{DD}$) -- in fact, even less than the average
statistical error of the sample $\sigma_E$.  In other words, though this fit is
accurate to only $20\%-30\%$ at any point (based on $10^{I_q}-1$ for
$q=1\ldots 4$), this accuracy is sufficient to apply constraints,
given the many orders of magnitude range spanned by $R(x)$.

\begin{deluxetable}{rllllll}
\tablecolumns{7}
\tablewidth{500pc}
\tablecaption{Behavior of Sample Fit as Order Increases}
\tablehead{
\colhead{ $q$} & \colhead{$\chi ^2$} & \colhead{$J$} & \colhead{$I$} & \colhead{$\sigma_{DF}$}  & \colhead{$\sigma_{DD}$} &\colhead{ $\sigma_E$} 
}
\startdata
 0 & 60.7 & 0. & 0.462 & 0.421 & 0.421 & 0.241 \\
 1 & 4.16 & 0.373 & 0.138 & 0.228 & 0.421 & 0.241 \\
 2 & 1.3 & 0.104 & 0.0813 & 0.203 & 0.421 & 0.241 \\
 3 & 1.61 & 0.0475 & 0.0929 & 0.198 & 0.421 & 0.241 \\
 4 & 2.4 & 0.105 & 0.143 & 0.182 & 0.421 & 0.241
\enddata
\label{tab:analogConvergence}
\end{deluxetable}

But while accurate fits \emph{exist}, the tools presented henceforth
provide few ways to identify them and
particularly estimate their error without resorting to knowledge of
the function being fit.   For example, as we increase
the available number of degrees of freedom, the root-mean-square difference
$\sigma_{DF}$
between the fit and the data decreases monotonically  to zero.   Even
at or near the optimal fit order, 
in our experience both  $\sigma_{DF}$ and $\sigma_E$, the expected
sampling-induced root-mean-square difference between data and
predictions [Eq. (\ref{eq:physical:diagnostic:sigmaBetweenFitAndData})],   correlate only
weakly with the true error in the fit ($I_q$).  A good
fit can be 
\emph{identified}%
\footnote{The best way to \emph{identify} a good fit, given sampled
  data, is with a  \emph{blind test}, where data points not involved
  in the fitting process are compared against the fit.  We have explored using
blind tests points and calculating $\chi^2_{q,b}$, the analagous
weighted squared difference between the fit and these test points.
In our test cases, $J_q$ provided just as much discriminating power
between fits as a blind test.
}
using the
weighted squared difference $\chi^2_q$, which does have a minimum
value  for a good fit.\footnote{In fact, $\chi^2_q$ also allows us to
  identify when the fit is as good as the sampling statistics allow,
  when  $\chi^2_q\simeq 1$} 
However, to understand and estimate the fit-induced error we must
compare fits of successive orders with 
\begin{eqnarray}
\label{eq:physical:diagnostic:Jq}
J_q &\equiv &\left[ 
  \int d^7x \left[ \log(\hat{R}_q/\hat{R}_{q-1})
 \right]^2 \right]^{1/2}
\end{eqnarray}

This comparison between fits of different orders
arises naturally from understanding how the error in the fit  $I_q$
reflects
a balance between (i) an increasing ability to fit $\log R$ exactly with more
basis polynomials and ever smaller ``truncation error'', decreasing
the magnitude $||P_{\perp q} \log R||$ for $P_{\perp q}$ the 
 and (ii) a
decreasing statistical accuracy with 
fewer independent degrees of freedom $N_s - N_{\le q}$ available
to constrain the $N_{\le q}$ independent coefficients in the
expansion.  These two terms contribute independently to the error at
any fit order, allowing us to estimate $I_q$ by an incoherent
superposition of ``truncation error'' and statistical error
\emph{without} explicitly calculating the best fit $\log\hat{R}_q$ itself:
\begin{eqnarray}
\label{eq:errorRelation}
I_q^2 &\simeq&||P_{\perp q} \log R||^2  
 +\sigma_E^2 \frac{N_{\le q}}{N_s-N_{\le q}} 
\end{eqnarray}
with 
\begin{eqnarray}
N_{\le q} &\equiv& \frac{(d+q)!}{d!q!}
\end{eqnarray}
Here $d=7$ is the dimension in which our functions are defined and
 $P_{\perp q}$  projects functions perpendicular to the space
of basis polynomials of order less than or equal to $q$
[cf. Eq. (\ref{eq:project:oned})]: for example,
if $\log R$ is a polynomial, $P_{\perp q}$ selects those coefficients
of $\log R$ of order less than q.  
In this and other cases, this expression very accurately
reproduces $I_q$; see for example Table  \ref{tab:analogConvergence}.

When sampling errors are relatively small (i.e., $n_\alpha \gg 1$, so
$\sigma_E\sim 0$), each successive approximation
order 
$\log \hat{R}_q \simeq P_q \log R$ is found by orthogonal projection
of $\log R$; therefore, for low fit orders the \emph{truncation error}
is just the magnitude of the next-highest-order correction to be
applied, $J_q$:  
\begin{equation*}
||P_{\perp q} \log R||\simeq ||\log \hat{R}_{q+1}-\log \hat{R}_{q}|| \equiv J_{q+1}
\end{equation*}
In other words, for low $q$ we expect $I_q\simeq J_{q+1}$, as
confirmed in Table    \ref{tab:analogConvergence}.

From our experience with this and other model problems, the true rms
fit error ($I_q$) is quite generally close to the change in the fit
to the next order ($J_{q+1}$) and from the previous order ($J_{q}$).
In particular, the difference between the current and
next-lowest-order fit ($J_q$) has a minimum similar in magnitude and
location to the minimum of $I_q$; see for example Table
\ref{tab:analogConvergence}.
We therefore use the order at which $J_q$ is smallest to select the
optimal fit order and its value as the characteristic
root-mean-square error in the fit.

\subsection{Volume estimation}
This error estimate provides the key tool needed to \emph{reliably and
  algorithmically surround} almost all of the points with 
merger rates $R$ consistent with constraints, using information only
about the fit, to the high level of accuracy needed to trust the
results of multiple constraints.  
For example,  observations of Galactic binary pulsars suggest double
neutron star binaries merge 
in the Milky Way merge at a rate  between $c_{\rm
  min}=\log(3\times 10^{-5}{\rm yr}^{-1})$ and $c_{\rm max}=\log(23\times
10^{-5}{\rm yr}^{-1})$.
This constraint range turns out to lie on the high end tail of what our model simulations produce
\citep{Chunglee-nsns-1,PSmoreconstraints}.
If our seven-dimensional model exactly described double neutron star
merger rates, then only a small fraction of model parameters
$\vec{x}\in V(C,\log R)$ ($5\%$ by
volume of the unit seven-dimensional cube, corresponding to
$|\vec{x}|<1.04$) reproduce observations.  However, even though these
merger rates correspond to the largest possible and therefore
best-sampled predictions from population synthesis,  the fit remains
sufficiently inaccurate so that only $1-r_+\simeq 18\%$ of all  constraint-satisfying
points would not be included in a prediction  $V(C,\log{\hat{R}})$
based on the fit alone.  
But based on Table
\ref{tab:analogConvergence}, we can estimate the characteristic
error in our fit by $J\simeq 0.05$.   Therefore,  when we compensate for this uncertainty
with a wider  constraint interval   $C^*\equiv C^*(J)$,
we much more accurately bound the set of constraint-satisfying points: 
$1-r_+(C^*,\log\hat{R})=5\%$.


Strictly, the fraction of constraint-satisfying points that will be
encompassed with a larger constraint interval $\tilde{C}$ depends strongly on
the width and placement of the original  interval $C$ -- intuitively,
an extremely narrow 
constraint requires extremely high-precision reconstruction
of $R(\vec{x})$, in turn possible only at high merger rates $R$,
where many events should be seen in simulation.  However, the case shown
here, with a very tight constraint (with only $5\%$ by volume of
parameters satisfying the constraint), is substantially more difficult
to satisfy than most individual observational constraints that can be
applied in practice.   Comparing our experience with model problems to the weak
constraints available, we are confident that fewer than $5\%$ of
constraint-satisfying systems will be omitted from $V(\tilde{C}, \log
\hat{R})$ when these volumes are constructed on the basis of real
population synthesis simulations and corresponding observations.
[The results of our broader array of model problems are available
online in \cite{PSutil2}.]

\section{Fitting: Population Synthesis Predictions}
\label{sec:ps:predictions}
Almost exactly the same fitting techniques presented above are
applied to our population synthesis data: we
perform a weighted least-squares fit of polynomial-like basis
functions over our
seven-dimensional space, for each of the event rates of interest 
[BHBH(m), BHNS(m), NSNS(m), NSNS(vw), WDNS(e), WDNS(m), SNIb/c, and
SNII, as introduced in \S\ref{sub:id}].    For each fit, we evaluate the fit quality of several
different polynomial orders to our data.  To minimize the possibility
of using more parameters than allowed by the number of  our data
points, we choose  as ``best'' fit  
that order which minimizes the relative difference between fit orders
$J_q$ (see below).
 However,
we know that the merger rate $R(x)$ must satisfy certain symmetries,
based on the manner in which we represent the kick velocity dispersion
as a sum of two maxwellians (i.e., we can switch labels associated
with the two distributions); therefore we limit the manner in which
those parameters associated with kicks can enter into the distribution
(see the Appendix).


\begin{deluxetable*}{l|rr|rrrrr|rr|r}[ht]
\tablecolumns{11}
\tablewidth{500pc}
\tablecaption{Statistics of population synthesis simulations for event
rates}
\tablehead{\colhead{Type}&\colhead{$N_s$}&\colhead{$N_{s,+}$}&
\colhead{$\bar{N}_{\le q}$}&\colhead{$q$}&\colhead{$\chi^{2}$}&\colhead{$\sigma_{DF}$}&\colhead{$J_q$}&%
\colhead{$\sigma_E$}&\colhead{$\sigma_{DD}$}&
\colhead{$\eta(10)$} 
}
\startdata
BH-BH(m) & 2930 & 1201 & 141 & 3 & 15.5  & 0.77 & 0.67 & 0.25 & 0.78 & 12\%   \\
BH-NS(m) & 2533 & 1334 & 141 & 3 & 11.1  & 0.53 & 0.40 & 0.23 & 0.71 & 2\%   \\
NS-NS(m) & 2803 & 2382 & 141 & 3 & 18.4  & 0.37 & 0.22 & 0.14 & 0.63 & 5\%  \\
NS-NS(vw)& 1325 & 1087 & 141 & 3 & 16.7  & 0.39 & 0.37 & 0.13 & 0.76 & 8\% \\
WD-NS(e) & 1770 & 1564 & 141 & 3 & 12.5  & 0.34 & 0.20 & 0.14 & 0.57 & 10\%  \\
WD-NS(m) & 1770 & 1658 & 141 & 3 & 16.4  & 0.34 & 0.19 & 0.13 & 0.45 & 11\%  \\
SN Ib/c   &1482 & 1482 & 141 & 3 &  7.8  & 0.07 & 0.06 & 0.02 & 0.11 & n/a   \\
SN II     &1482 & 1482 & 141 & 3 &  6.0  & 0.04 & 0.02 & 0.03 & 0.17 & n/a  
\enddata
\label{tab:SimulationsAndFits}
\end{deluxetable*}

Table \ref{tab:SimulationsAndFits} summarizes the properties of the least-squares fits
we applied to our archived population synthesis results.
The first column provides a brief label for the fit, as described in
greater detail in the text.  The next two columns summarize the amount
of available information contained in our population synthesis
archive: $N$ is the number of population synthesis models with
unbiased data (i.e., where all plausible progenitors for the target
event have been included), whereas $N_+$ is a smaller number of
models with unbiased data containing one or more events (i.e., for
which an estimate of the rate, rather than merely an upper bound, is
possible).   
The next block of five columns describes properties and diagnostics of a
weighted least-squares fit applied to our data.  The first two
columns, $q$ and $\bar{N}_{\le q}$, merely indicate the polynomial order and
number of degrees of freedom involved in our fit ($\bar{N}_{\le q}$
differs from $N_{\le q}$ introduced earlier due to the symmetry requirement
described above).   In all the cases
shown here, the optimal polynomial order produces far fewer degrees of
freedom than population synthesis simulations (i.e., $\bar{N}_{\le q}
\ll N_{s,+}$).  

The next  three columns provide critical
diagnostics of our fit: $\chi^2$, $\sigma_{DF}$, and $J_q$ 
[Eqs. (\ref{eq:physical:diagnostic:chi2q},\ref{eq:physical:diagnostic:sigmaBetweenFitAndData},\ref{eq:physical:diagnostic:Jq})].
All three columns roughly measure our ``goodness-of-fit''; 
based on our experience with model problems and our understanding of
the values these quantities should take for fits dominated by
statistical and truncation errors,
all three
universally indicate that truncation error dominates -- that is, that our low-order polynomial basis functions are  not sufficiently
general models to match the exact merger rate $R(x)$ implied by our
population synthesis simulations.
For example, the limiting value of $J_q$ is
significantly above the level of sampling error
$\sigma_E/\sqrt{N_{s,+}/\bar{N}_{\le q}}$ suggested by the second term in
Eq. (\ref{eq:errorRelation}) 
-- that is, above the level of error
expected when averaging $N_{s,+}/N_{\le q}$ simulation samples per
degree of freedom, each with characteristic error $\sim \sigma_E$.
This high level of error suggests fit accuracy is limited by
truncation error -- inability to fit the exact form of
$R(x)$ with our basis polynomials -- rather than sampling uncertainties at each point.\footnote{Additionally, when fitting supernovae rates as a
  function of population synthesis parameters for \emph{single} stars,
  where only one of our parameters (wind strength) enters, we find the
  rate has  a moderately complex functional form that requires high-degree polynomials
  to fit.  We expect similarly complex behavior in the
  multidimensional case, and interpret the large $\chi^2$
  correspondingly.}  
Additionally, as illustrated in our discussion of seven-dimensional
model problems, the latter two columns provide the \emph{logarithmic
  uncertainty} in our fits; for example, fits with  $J\simeq 0.33$
are known to within a factor two with one-sigma confidence. 
These columns should be contrasted with the next two, which provide
the average sampling-induced uncertainty  $\sigma_E$ in the data
[Eq. (\ref{eq:physical:diagnostic:sigmaBetweenFitAndData})] and the
characteristic range of merger rates $\sigma_{DD}$ seen in simulation
[Eq. (\ref{eq:physical:diagnostic:sigmaOfData})].
Our fits remain useful so long as their uncertainties are much less
than the range of $\log R$ (i.e., so long as $J\lesssim\sigma_{DD}$).  
As seen in Table \ref{tab:SimulationsAndFits}, our best fit satisfies
this requirement
($J\lesssim \sigma_{DD}$)  for each populations of interest.

The last
column provides the fraction $\eta$ of simulations with \emph{no}
events when more than $10$ would be expected based on the fit.  Based
on an \emph{average} uncertainty of a characteristic factor of two in
the fit (our best fits have $J\simeq 0.33$, corresponding to a merger
rate known to within a factor $2\simeq 10^{J}$),
 we would estimate
that the fraction of simulations $\eta$ which produce no merger events
when more than 10 should be seen based on our \emph{uncertain
  estimate} of the event rate should be comparable to or smaller than the
fraction of simulations which should produce $10/2=5$ events but in
fact produce zero, namely $\eta \lesssim p(0,10/2)\simeq 0.7\%$ [Eq. (\ref{eq:def:poisson})].
However, regions with the lowest
merger rates will be undersampled and therefore have characteristic
uncertainties marginally larger (e.g., up to a factor $3$ to $4$),
leading to $\eta$ of a few to ten percent.


\optional{
\subsection{Sample Completeness}
Some population synthesis archives which provide unbiased data for
event rate estimates have zero events (i.e., the number of population
synthesis models 
$N_s$  is more than the number of models with at least one observed
event $N_{s+}$, in
Table~\ref{tab:SimulationsAndFits}).  Because of the enormous range of
population synthesis sample sizes $N$ used, sometimes zero events for
rare events should be expected [i.e., merging BH-BH binaries:
BHBH(m)].  On the other hand, a significant set of large-$N$
population synthesis archives with $n=0$ for a relatively
\emph{common} event might imply that the event has been systematically
undersampled: in some region, the underlying event rate might be significantly
lower than our archive collection has been able to probe.  Consequently these low event rates are not represented in our database of rates to be fit.

To distinguish between the two possibilities in cases where a large number of
zero-event samples ($N_s-N_{s+}$) indicate that undersampling is
possible [i.e.,  BHBH(m) and NSNS(vw)], we scatter-plot $\log_{10}N_s$
versus $\log_{10}n$, superimposed by the line $n/N=R_{99\%}/s$,
where $R_{99\%}$ is the 99\% lower limit predicted on the basis of the
fit (see
Figure~\ref{fig:popsyn:completeness}).  If the fit is of high quality , very few systems should lie below this
line; further, if our sample is fairly complete, \emph{very} few high-$N$
systems should lie below this line.    We see no compelling indication
of undersampling.\footnote{Unfortunately, the hardest simulations had to
be run the longest, typically producing an over-represented collection
of low-$n$, high-$N$ simulations.  As a result, the distribution
of $N$ and $n$ is often biased at the largest $N$.}
}

\subsection{Results}
\noindent \emph{Supenovae}:
Given our choice of stellar mass interval probed in our simulation (i.e., $m_1>4 M_\odot$), supernovae
occur extremely
frequently, providing us with superb statistics at low cost.
However, our limited set of basis functions can only with difficulty
reproduce the observed variation in supernovae rates: even though
 SN rates for models in our archive are at times known to $1\%$ (i.e.,
involving $10^4$ or more sampled events), or $0.004$ in the log, our
optimal fit differs significantly from the data, by
$\sigma\simeq O(0.04)$ in the log.   


Nonetheless, as discussed in the forthcoming
\citet{PSmoreconstraints}, the supernovae rate remains a striking success of the
\emph{StarTrack} population synthesis code and our normalization
conventions (e.g., $\dot{M}\simeq 3.5 M_\odot {\rm yr}^{-1}$).   No matter what combination
of population synthesis parameters we choose, the predicted SN rates
lie well within the observational constraints found by
\citet{Cappellaro:SNa}.  

\noindent \emph{WD-NS  binaries}:
As with supernovae, white dwarf-neutron star binaries occur fairly
frequently,
allowing us to
accumulate fairly good statistics over a broad range of population
synthesis parameters.  Additionally, based on  the distribution of $N_s$
versus $n$ (i.e., as in Figure~\ref{fig:popsyn:completeness}), our
sample shows no signs of systematic incompleteness: we appear to have
covered the full range.    Though our polynomial fits continue to
introduce systematic error, the resulting fit behaves well throughout
the range.

\noindent \emph{BH-BH binaries}: 
Double black hole binaries, in contrast, occur extremely
infrequently,
especially with an assumed maximum NS mass of $2.5 M_\odot$, instead
of $2.0 M_\odot$, as assumed in \citet{PSutility,PSconstraints}.  [The
inefficient formation of coalescing BH-BH binaries is discussed and
explained in more detail by \cite{ChrisBH2007}.]
Nonetheless, by using special-purpose partitions, 
\citet{PSutility} accumulated a large ($\simeq 500$ simulations) sample with good ($n\simeq 10$)
statistics on a large subset of parameter space; 
though these simulations assumed a maximum NS mass of $2 M_\odot$, we
\emph{post facto} changed the maximum NS mass to $2.5 M_\odot$.  By adjoining the
results of
general-purpose simulations \emph{not} assured of good statistics, this sample has
since been enlarged by a factor $\simeq 5$, with emphasis on the same subset of parameter
space presented in \citet{PSutility} (see
Figure~\ref{fig:popsyn:kicksampling}).    Finally, though the
distribution of $n$ versus $N_s$
(Figure~\ref{fig:popsyn:completeness}) suggests the lowest BH-BH
merger rates may not be very well-resolved, we have no reason to
suspect we have any significant under-resolved region: simulations larger than $10^6$
binaries consistently produce several merger events ($n\gg 0$). 

Nonetheless, 
the BHBH(m) fit is
comparatively poor: even though the average simulation with binary black holes
produces enough to fairly accurately determine the rate
(i.e., $\sigma_E\simeq 0.25$), our best fit is barely more
accurate than  approximating the average BH-BH merger rate seen in simulations
(i.e., compare the characteristic fit error $J_q\simeq 0.67$ with the range of
BH-BH merger rates seen in simulation, $\sigma_{DD}\simeq
0.78$).

\noindent\emph{Results: NS-NS and BH-NS binaries}:
Despite (or perhaps because of) a concerted effort to accumulate good
statistics targeted  specifically to these classes, fits  for NS-NS
and BH-NS event rates are statisticallly implausible as measured by
$\chi^2$.   Judging from Figure~\ref{fig:popsyn:completeness}, the
NSNS(m) and BH-NS(m) are comparatively well-resolved: longer
simulations fairly consistently have a lower chance of producing $n=0$
results.    For this reason, we strongly 
suspect some feature of these underlying rate functions are
poorly-described by our basis functions; we intend to more thoroughly
test this hypothesis (with better statistics) by comparing these fits to
 nonparametric estimates in a future paper.  

\optional{
\textbf{I think this is all very old info...must review.}
Again judging from Figure~\ref{fig:popsyn:completeness}, the
sparsely-sampled NSNS(vw) rate is likely undersampled: the longest
simulations, oddly, are \emph{more} likely than usual to produce only
zero or one event!  Though we are at present unable to directly characterize the region
responsible, we suspect that certain combinations of population
synthesis parameters produce an abnormally low probability for
recycling pulsars in wide NS-NS binaries. 
}




\section{Conclusions}
To  develop a more comprehensive understanding of population synthesis
predictions  and   to allow
those theoretical predictions to be systematically compared with
observations of the end products of high-mass single and binary
stellar evolution, we have fit eight predictions from 
the \emph{StarTrack} code over seven of its input parameters.  
These fits are available on requests sent to the first author.  In a companion
paper, \citet{PSmoreconstraints} apply these fits along with 
estimates of their systematic errors to discover robust constraints on
the seven parameters that enter into population synthesis.  Additionally, we have 
demonstrated that in analagous model problems the
constraint-satisfying region defined by using these fits can, under
appropriate conditions, very accurately trace the underlying
constraint-satisfying region.
Finally, we have presented a thorough diagnostic formalism, including
a large list of diagnostic quantities and tests
($I,J,\sigma_{E,DF,DD},\chi^2,\eta, r_+$, etc.), which can be applied to
studies of fits to large archives of \emph{any} (individual) population
synthesis code.

\acknowledgements

We thank NCSA for providing us with
resources used to perform many of the computations in this text.
This work is partially supported by an NSF Gravitational
Physics grant PHYS-0353111, a David and Lucile
Packard Foundation Fellowship in Science and Engineering, and a
Cottrell Scholar Award from the Research Corporation to VK.  KB
acknowledges support from the Polish Science Foundation (KBN) Grant
1P03D02228.
VK is grateful for the hospitality and partial financial support of the KITP, UCSB for the period of June and July 2006.

\bibliography{%
popsyn,popsyn_gw-merger-rates,%
observations-pulsars,observations-pulsars-kicks,observations-supernovae,%
gw-astronomy-mergers,%
mm-general,mm-statistics,%
short-grb,Astrophysics}

\appendix
\section{Sampling, prior distributions and the likelihood of results}
\label{ap:sampling}
In Figure \ref{fig:popsyn:distributions} we compare the
distribution of merger rates  derived directly from our current
database of  simulations (dotted lines) to the distribution
of merger rates derived from the corresponding fit (solid lines) obtained using
the methodology described here.   Unlike 
Figure 3 of \cite{PSconstraints}, in which the simulations being fit
had parameters drawn randomly from the \emph{entire} range of
parameters allowed, in this study the simulations were \emph{not}
randomly distributed through the entire parameter range; see for
example Figure \ref{fig:popsyn:kicksampling}.  In effect, the
rate distributions drawn from simulations and fits assume
significantly different priors 
for population synthesis model parameters, the former favoring lower
kicks.  Therefore, although the distributions \emph{appear} different
in shape,
they reflect the same physical process, just with different
priors regarding what model parameters are likely.
In particular,  the  \emph{fitted}
  rate results 
represent the unbiased distributions of rates with uniform coverage of
the seven-dimensional model parameter space. Furthermore these
distributions fully account for uncertainties in the fits, as
described in \S\ref{sec:ps:predictions}.


To phrase the same statement more abstractly, our prior
expectations about the relative likelihood of population synthesis
parameters (e.g., kicks) influences our expectations for the relative
likelihood of different merger rates.  
If at any point our knowledge of binary properties, evolution, and
compact object kicks improves to a degree such that we can
confidently move away from flat priors into favoring certain, more
specific priors for the model parameters, then the shape of the
probability distributions derived from fits will change accordingly,
expressing the influence of the adopted priors.  

\begin{figure}
 \includegraphics[width=500pt]{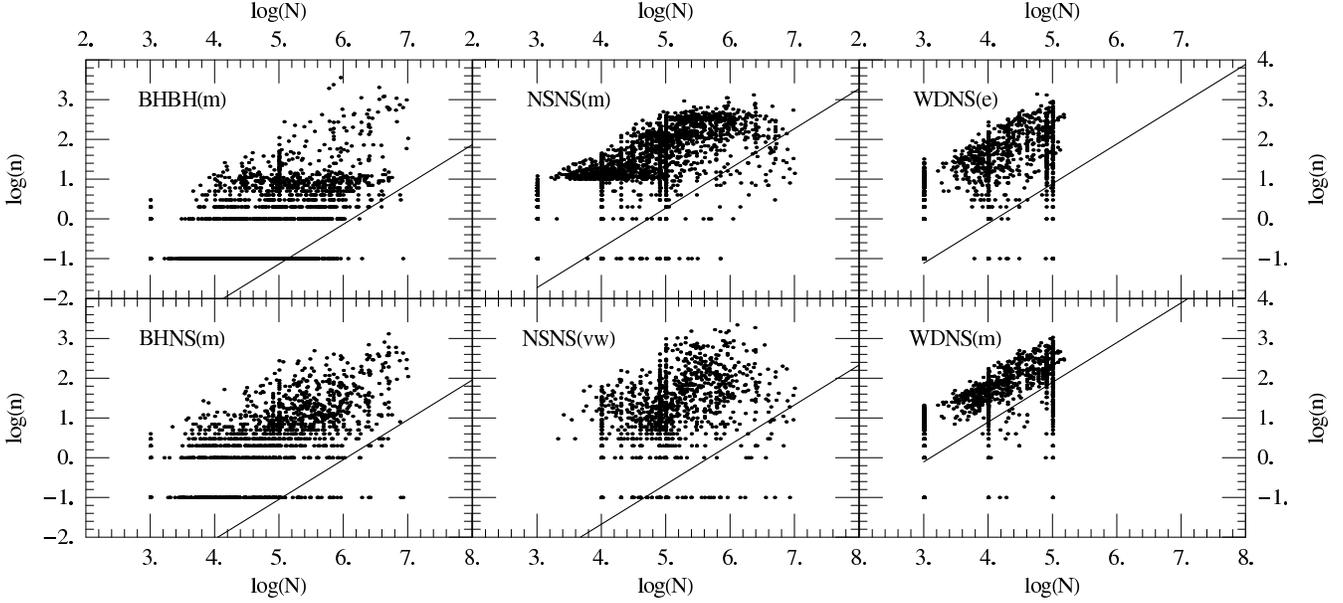}
\caption{\label{fig:popsyn:completeness} 
Log-log scatter plot of the number of events seen in a sample against
the population synthesis sample size.   Also shown  is a
diagonal line corresponding to the 99\% lower bound on the
fit-predicted rate distribution, translated from physical rate into
expected number of events seen per unit population synthesis sample
size.  For completeness, models with zero events are shown with
$\log_{10}n$ as $-1$.
}
\end{figure}

\begin{figure}
 \includegraphics[width=500pt]{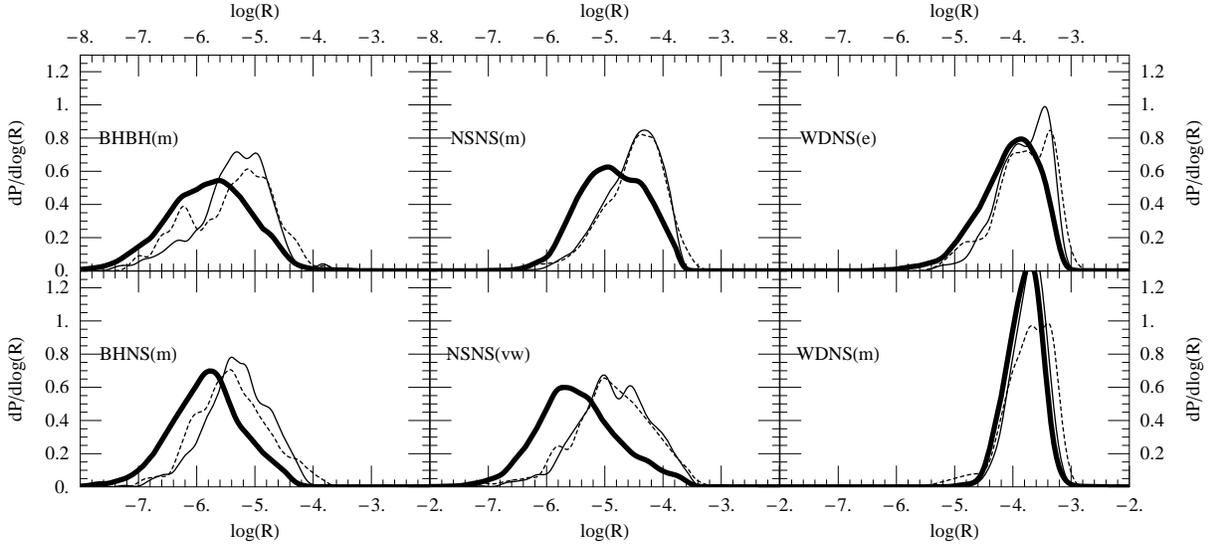}
\caption{\label{fig:popsyn:distributions} 
Probability distributions for formation rates for different types of
binaries in the Milky Way.
Dotted lines:
the distributions of  rates
derived {\em directly} from our database of simulations, which does
{\em not regularly sample} the model parameter space; cf. Figure
\ref{fig:popsyn:kicksampling}.
Thin solid curves: the distribution of
different formation rates derived \emph{using fits}, using precisely
the same sampling as the raw data (dotted line).
Heavy solid curve: the distributions derived from our fits to the
simulations when the model parameter space is sampled uniformly.
In all cases, the formation rate data is smoothed with a gaussian of
standard deviation $0.1$.  
It is evident that: (i) the quality of the fits is rather good; and
(ii) the change in the shape of the distributions is primarily because
of the different priors between the dotted and thick-solid lines. Last
we note that: (i) the rate results for binaries involving BHs are most
uncertain because of the assumed maximum NS mass of $2.5 M_\odot$  that
reduces the overall formation rates for BH binaries (compared to an
assumed maximum mass of $2 M_\odot$); (ii) all rates quoted in this studies
are for Milky-Way like galaxies and we do not account for any
differences associated with elliptical galaxies.  
}
\end{figure}

\section{Polynomial basis functions}
\label{ap:poly}
To parameterize supernovae kick distributions, \emph{StarTrack}
employs three parameters, $\sigma_1$, $\sigma_2$, and $s$, which represent the
superposition of two Maxwellian kick distributions with probabilities
$s$ and $1-s$.  The physical predictions associated with $(\sigma_1,\sigma_2,s)$
are therefore identical to those of $(\sigma_2,\sigma_1,1-s)$.  To improve the
physical significance of our fit, we have chosen to employ basis
polynomials which \emph{enforce} this requirement to all orders.

Specifically, rather than allow for homogeneous basis polynomials in
these parameters, we use the following, for arbitrary $p$ and $q$:
\begin{eqnarray}
\sigma_1^p s^q + \sigma_2^p (1-s)^q\\
2 \sigma_1^p \sigma_2^ps (1-s) + \sigma_1^{2p} s^2 + \sigma_2^{2p}(1-s)^2
\end{eqnarray}
Because $s$ must enters in a heterogeneous manner to preserve our
desired symmetry, these basis polynomials are of a fixed order in all
kick parameters.  For the purposes of order counting when constructing
fits, the first polynomial is denoted order $p+q$ and the second order
$p$.




\end{document}